\documentclass[10pt,conference]{IEEEtran}
\IEEEoverridecommandlockouts
\usepackage{cite}
\usepackage{amsmath,amssymb,amsfonts}
\usepackage{algorithmic}
\usepackage{graphicx}
\usepackage{subcaption}
\usepackage{textcomp}
\usepackage{pifont}
\usepackage{caption}
\usepackage{xcolor}
\usepackage{colortbl}
\usepackage{enumitem}
\usepackage{multirow}
\usepackage{url}
\usepackage{hyperref}
\usepackage{balance}
\usepackage[linesnumbered,ruled,vlined]{algorithm2e}
\usepackage{amsmath,amssymb} 
\usepackage{changepage}
\usepackage[markup=underlined]{changes} 

\usepackage{booktabs}
\usepackage{siunitx}

\usepackage{graphicx}   
\usepackage{booktabs}   
\usepackage{makecell}   
\usepackage{array}      

\usepackage{setspace}

\usepackage[most, breakable]{tcolorbox}
\def\BibTeX{{\rm B\kern-.05em{\sc i\kern-.025em b}\kern-.08em
    T\kern-.1667em\lower.7ex\hbox{E}\kern-.125emX}}

\usepackage{colortbl}

\usepackage{arydshln}

\definecolor{green1}{RGB}{240, 250, 240} 
\definecolor{green2}{RGB}{200, 230, 200} 
\definecolor{green3}{RGB}{150, 210, 150} 
\definecolor{green4}{RGB}{100, 190, 100} 
\definecolor{green5}{RGB}{50, 160, 50}

\usepackage{xspace}
\newcommand{\mypara}[1]{\smallskip \noindent\textbf{#1.} \xspace}
\usepackage{longfbox}
\definecolor{ashgrey}{rgb}{0.7, 0.75, 0.71}
\definecolor{grey}{rgb}{0.6,0.6,0.6}

\usepackage{cleveref}                 

\usepackage{subcaption}      
\begin{document}


\title{SemGuard: Real-Time Semantic Evaluator for Correcting LLM-Generated Code}


\author{
  \IEEEauthorblockN{Qinglin Wang}
  \IEEEauthorblockA{Shandong Normal \\ University \\
  Jinan, China \\
  2023317094@stu.sdnu.edu.cn}
  \and
  \IEEEauthorblockN{Zhihong Sun}
  \IEEEauthorblockA{Shandong Normal \\ University \\
  Jinan, China \\
  2022021002@stu.sdnu.edu.cn}
  \and
  \IEEEauthorblockN{Ruyun Wang}
  \IEEEauthorblockA{Institute of Information Engineering, \\ Chinese Academy of Sciences \\
  Beijing, China \\
  wangruyun@iie.ac.cn}
  \and
  \IEEEauthorblockN{Tao Huang}
  \IEEEauthorblockA{Shandong Normal \\ University \\
  Jinan, China \\
  2022317095@stu.sdnu.edu.cn}
  \and
  \IEEEauthorblockN{Zhi Jin}
  \IEEEauthorblockA{Key Lab of HCST (PKU), \\ MOE; SCS \\
  Beijing, China \\
  zhijin@pku.edu.cn}
  \and
  \IEEEauthorblockN{Ge Li}
  \IEEEauthorblockA{Key Lab of HCST (PKU), \\ MOE; SCS \\
  Beijing, China \\
  lige@pku.edu.cn}
  \and
  \IEEEauthorblockN{Chen Lyu\IEEEauthorrefmark{1}}
  \IEEEauthorblockA{Shandong Normal \\ University \\
  Jinan, China \\
  lvchen@sdnu.edu.cn}\thanks{\IEEEauthorrefmark{1}Corresponding author.}
}

\maketitle

\begin{abstract}

\textit{Large Language Models} (LLMs) can translate natural language requirements into code, yet empirical analyses of representative models reveal that \emph{semantic errors}—programs that compile but behave incorrectly—constitute the majority of observed faults (e.g., $>$60\% on DeepSeek-Coder-6.7B and QwenCoder-7B). Post-hoc repair pipelines detect such faults only \emph{after} execution, incurring latency, relying on incomplete test suites, and often mis-localizing the defect. Since semantic drift originates in the autoregressive decoding process, \emph{intervening while the code is being generated} is a direct way to stop error propagation.  Constrained-decoding approaches such as ROCODE attempt this, but still wait until the entire program runs to obtain feedback and use entropy heuristics that do not truly capture semantics.  A more effective solution must inject \emph{semantic} signals—early and precisely—into the decoding process.We present \textbf{SemGuard}, a semantic-evaluator-driven framework that performs real-time, line-level semantic supervision.  To train the evaluator, we build \textit{SemDiff}, the first dataset with fine-grained annotations that mark the exact line where a correct and an incorrect implementation diverge.  The evaluator, once embedded in the LLM’s decoder, flags deviations on partial code, rolls back to the faulty line, and guides regeneration—without executing the program or requiring test cases. Across four benchmarks, SemGuard consistently outperforms state-of-the-art baselines.  It lowers the semantic error rate by \textbf{19.86\%} on \textit{SemDiff} relative to ROCODE, and lifts Pass@1 by \textbf{48.92\%} on the real-world \textit{LiveCodeBench} with CodeLlama-7B.  Similar gains hold for StarCoder2-7B on \textit{MBPP} and for DeepSeekCoder-6.7B on the Java benchmark \textit{SemDiff-Java}, demonstrating model- and language-agnostic effectiveness.
\end{abstract}

\section{Introduction}
\label{Intro}

\begin{figure}
    \centering
    \includegraphics[width=\linewidth]{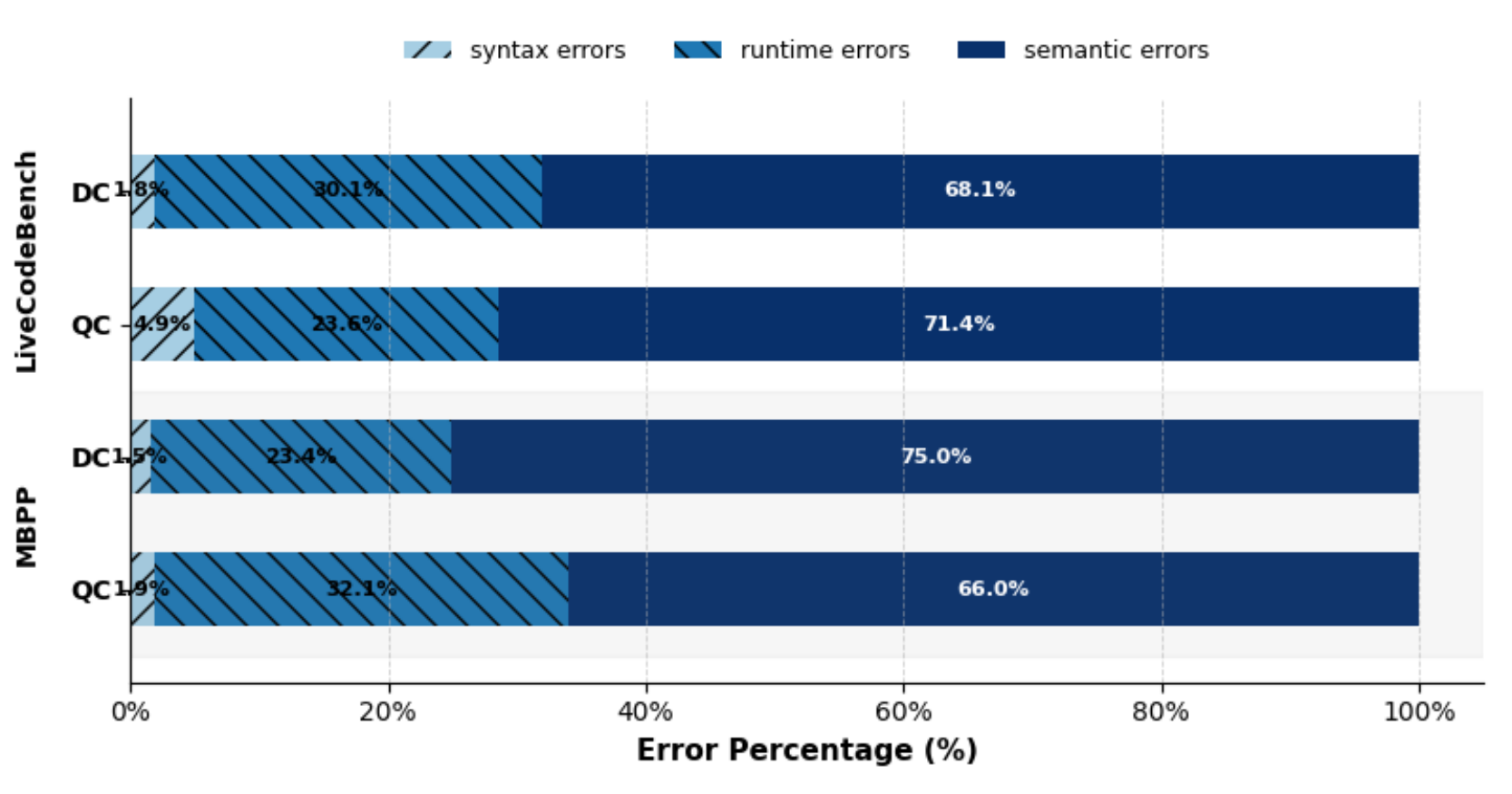}
    \caption{
Distribution of error types in LLM-generated code.  
    Results are collected using greedy decoding with DeepSeek-Coder-6.7B (\textbf{DC}) and Qwen-Coder-7B (\textbf{QC}) on the \textit{MBPP} and \textit{LiveCodeBench} benchmarks.
    }
    \label{fig:预实验}
\end{figure}

\textit{Large Language Models} (LLMs), exemplified by GPT-o3~\cite{openai2024gpto3} and DeepSeek-R1~\cite{guo2025deepseek}, have significantly advanced automated software engineering tasks by translating natural language into executable code~\cite{ghaemi2024transformers}. However, the generated programs frequently contain subtle yet critical \textit{semantic errors}, where code compiles successfully but deviates from intended functionality. As shown in Figure~\ref{fig:预实验}, semantic errors account for over 60\% of faults in LLM-generated code, surpassing syntax and runtime errors combined. This prevalence arises primarily because existing debugging tools (compilers, static analyzers, and unit tests) lack sufficient semantic awareness, and the autoregressive decoding process of LLMs propagates initial semantic deviations throughout subsequent code generation, exacerbating early mistakes.

Previous efforts, notably ROCODE~\cite{jiang2024rocode}, have attempted to mitigate these semantic inaccuracies by integrating traditional program analysis tools within the decoding phase of LLMs. Although ROCODE effectively curbs syntactic and runtime errors, it remains limited in addressing semantic errors due to two critical drawbacks: \textbf{\ding{182} Post-generation semantic detection.} ROCODE verifies semantic correctness only after the complete program has been generated and corresponding test cases executed, resulting in delayed identification and rectification of semantic errors. This post-generation validation not only reduces development efficiency but also introduces potential security risks by running potentially erroneous and unverified code~\cite{he2023large}. \textbf{\ding{183} Imprecise backtracking-point localization.} ROCODE employs entropy-based heuristics to pinpoint the origins of semantic deviations. However, entropy inherently reflects prediction uncertainty rather than causal relationships, leading to inaccurate identification of the true sources of errors. Consequently, substantial segments of correctly generated code may be incorrectly discarded during backtracking, causing unnecessary computational overhead and diminished generation quality.

To address these limitations, we integrate a lightweight \textbf{semantic evaluator} into the decoding process, allowing real-time assessment of semantic correctness in \emph{partial code}. By identifying semantic deviations before code execution, our approach immediately backtracks to the erroneous line and regenerates code, effectively curbing error propagation. However, the primary difficulty lies in \textbf{training a dependable semantic evaluator}: unlike syntax errors, semantic correctness is implicit and highly context-dependent, complicating large-scale formalization, detection, and annotation.

Specifically, we encounter two core challenges:
\textbf{\ding{182} Scarcity of line-level semantic annotations.} Existing benchmarks such as \textit{APPS}~\cite{hendrycks2measuring}, \textit{CodeContests}~\cite{li2022competition}, and \textit{CodeNet}~\cite{puri2021codenet} rarely indicate exactly \emph{where} semantic deviations occur or precisely \emph{how} the erroneous code diverges from correct solutions.
\textbf{\ding{183} Undecidable semantic validity of partial code.} Without full program context or reference specifications, even experts find it challenging to reliably determine semantic correctness for isolated code fragments. Manual annotation is thus slow and error-prone, while automated error-localization methods—such as spectrum-based fault localization (e.g., Tarantula~\cite{jones2005empirical}, Ochiai~\cite{ochiai2012photoelectrochemical}, Dstar~\cite{wong2013dstar})—are ineffective, as generated partial fragments often lack complete execution traces for meaningful analysis.



To overcome these obstacles, we devise a dedicated data pipeline to generate fine-grained annotations of semantic deviations at the code-fragment level. Specifically, we extract pairs of highly similar correct and incorrect implementations from the \textit{CodeNet} dataset, identifying precise lines where semantic deviation begins. For code pairs with minimal differences, segmentation occurs directly at the differing line; for more complex cases, we utilize LLMs to semi-automatically pinpoint semantic deviation points, enabling scalable and high-quality annotation.

Leveraging this pipeline, we construct \textit{SemDiff}, a dataset providing fine-grained annotations of semantic deviations. Using \textit{SemDiff}, we train a semantic evaluator that judges partial code correctness in real time. This evaluator is embedded into the decoding process of an LLM, monitoring semantic validity of intermediate outputs continuously. Upon detecting a deviation, the LLM promptly backtracks and regenerates from the faulty line, preventing error propagation. We term this framework \textbf{SemGuard}, a \emph{semantic-evaluator-driven} code generation system combining real-time semantic detection and targeted backtracking to proactively safeguard logical correctness in LLM-generated code.

We extensively evaluate SemGuard across four benchmarks—\textit{SemDiff}, \textit{MBPP}, \textit{LiveCodeBench}, and \textit{SemDiff-Java}—to verify its effectiveness, generalizability, and cross-language adaptability. Experimental results consistently show that SemGuard substantially outperforms state-of-the-art baselines in reducing semantic errors. Specifically, on \textit{SemDiff}, SemGuard achieves a \textbf{19.86\%} lower semantic error rate compared to ROCODE, without requiring code execution or test cases. On Python tasks, SemGuard boosts the Pass@1 scores significantly: by \textbf{11.14\%} with StarCoder2-7B on \textit{MBPP} and by \textbf{48.92\%} with CodeLlama-7B on the challenging \textit{LiveCodeBench}. SemGuard also demonstrates strong language-independent performance, improving DeepSeekCoder-6.7B’s Pass@1 by \textbf{25.09\%} on \textit{SemDiff-Java}. These outcomes confirm SemGuard's robust applicability across diverse models and programming languages.





In summary, our work makes three key contributions:

\setlist[itemize]{left=0pt}
\begin{itemize}
    \item \textbf{SemGuard: Real-time Semantic Detection and Intervention.}  
    We propose SemGuard, a lightweight semantic evaluator embedded within LLM decoding, enabling immediate line-level semantic checking and targeted backtracking. This effectively prevents semantic errors from propagating, preserving previously generated correct code.

    \item \textbf{SemDiff: Automated Semantic Annotation Pipeline.}  
    We introduce \textit{SemDiff}, the first dataset providing precise line-level semantic annotations. Built via a diff-guided and LLM-assisted pipeline, \textit{SemDiff} pairs semantically correct and incorrect code fragments, pinpointing exact semantic deviation points to facilitate training robust semantic evaluators.

    \item \textbf{Cross-language and Cross-model Validation.}
 We conduct comprehensive evaluations using four representative LLMs—StarCoder2, CodeLlama, DeepSeek-Coder, and QwenCoder—on diverse benchmarks covering both Python and Java tasks. The consistent performance improvements across languages and models demonstrate SemGuard’s broad applicability and cross-language generalizability.
 For reproducibility and future research, we publicly release the \textit{SemDiff} dataset, trained evaluators, and the complete, open-source implementation of SemGuard.\footnote{https://github.com/wwwql/SemGuard}
\end{itemize}

\begin{figure*}
    \centering
    \includegraphics[width=\linewidth]{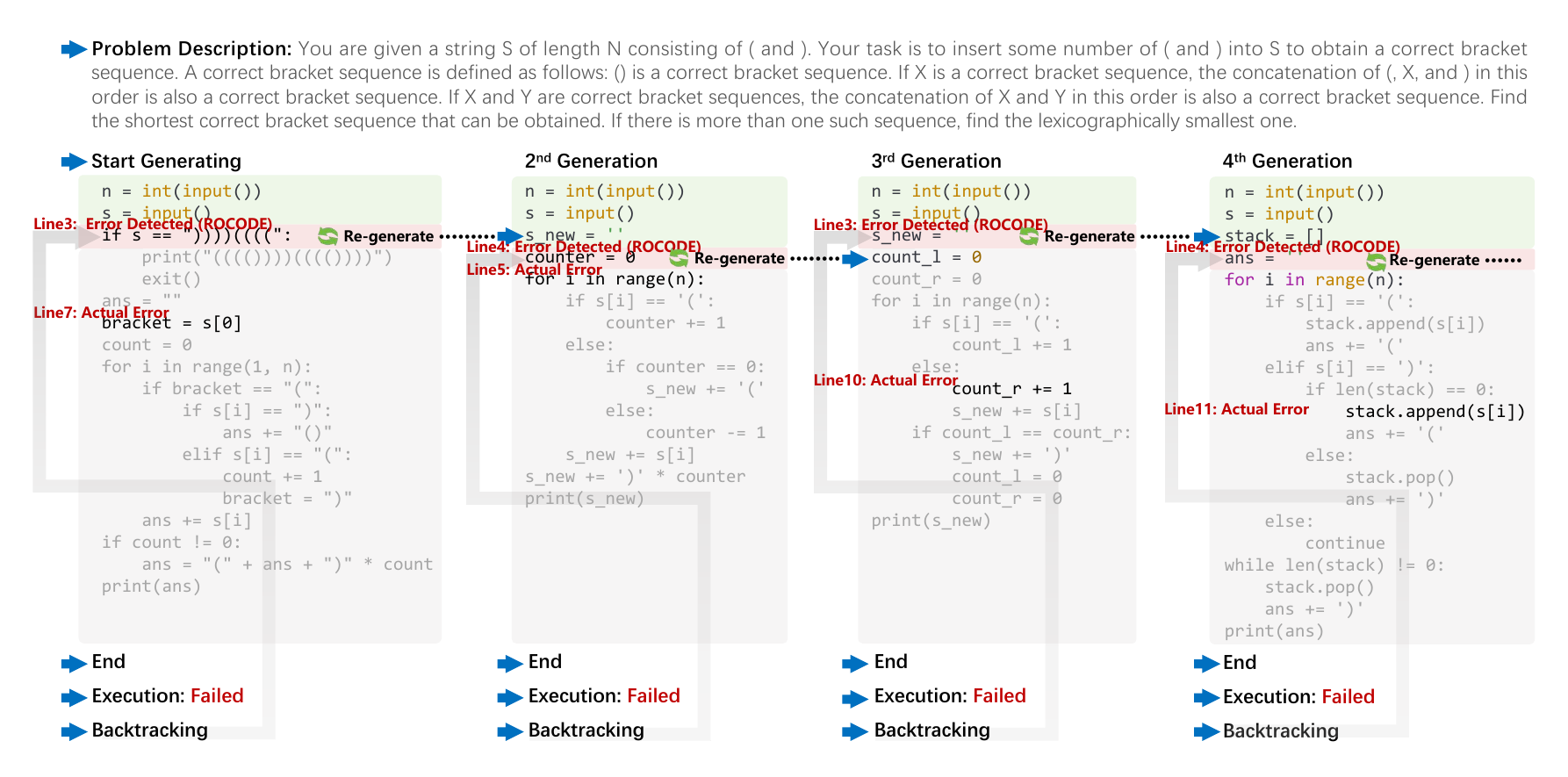}
    \caption{ROCODE backtracking sequence (DeepSeekCoder-6.7B) on CodeNet task \#03696.  }
    \label{fig:motivating}
    \vspace{-2mm}
\end{figure*}

\section{Motivation}

\subsection{A Concrete Example: Limitations of ROCODE}

ROCDOE reduces the occurrence of syntax and runtime errors by introducing a backtracking mechanism based on program analysis and test cases to locate decoding deviations. However, its design reveals certain limitations when dealing with semantic errors caused by semantic deviation during the decoding process.

As shown in Figure~\ref{fig:motivating}, we examine a representative task from CodeNet that asks the model to transform a string of parentheses into a balanced form using the fewest insertions. Among multiple valid answers, the lexicographically smallest one should be returned. ROCODE first generates an initial program, then applies three rounds of backtracking, but still fails to produce the correct result. 

The root cause lies in an early semantic error: during the first generation, the variable assignment \texttt{bracket = s[0]} captures only the first character’s bracket type. Subsequent logic relies on this local reference rather than tracking the actual number of unmatched left parentheses, which is crucial for correct insertion logic. This subtle semantic deviation gradually shifts the model's reasoning trajectory, leading to accumulation of errors.

Despite backtracking, ROCODE cannot effectively recover. This failure reflects two deeper limitations:

\begin{itemize}
  \item \textbf{Delayed semantic detection.} ROCODE performs semantic validation only after the full program is generated and executed on test cases. This example underwent four complete rounds of code generation, with semantic errors detected after execution in each round. This delay not only wastes computation but also poses security risks by executing unverified code.
  
  \item \textbf{Inaccurate rollback localization.} ROCODE uses entropy-based heuristics to choose rollback points, but entropy reflects uncertainty—not semantic deviation. For instance, during the three backtracking processes, ROCODE repeatedly locates the backtracking point incorrectly at the beginning of the code, rather than at the actual location where the semantic deviation occurs. Especially during the process from the third generation to the fourth generation, the true semantic error appears at line 10 (\texttt{count\_r +=}), but the model rolls back to line 3 (\texttt{s\_new = ''}). This misidentification truncates correct code and causes large-scale, unnecessary regeneration.
\end{itemize}

\subsection{Key Idea: Real-Time, Granular Semantic Supervision}



This example underscores that post-hoc test-based correction alone is insufficient; semantic correctness must instead be proactively ensured \emph{during} code generation. To achieve this, we propose integrating a lightweight \textbf{semantic evaluator} directly within the LLM decoding process, enabling real-time, fine-grained semantic checking. The evaluator continuously analyzes partial programs as they are generated, immediately detects semantic deviations, and initiates precise, line-level rollback and regeneration. This approach effectively halts error propagation with minimal disruption to previously correct code fragments.

Implementing such an evaluator is non-trivial because fine-grained semantic labels are scarce and partial-code semantics are hard to judge—challenges already detailed in Section \ref{Intro}. In the next section we explain how the proposed SemGuard framework overcomes these obstacles.

\section{Approach}

\begin{figure*}
    \centering
    \includegraphics[width=\linewidth]{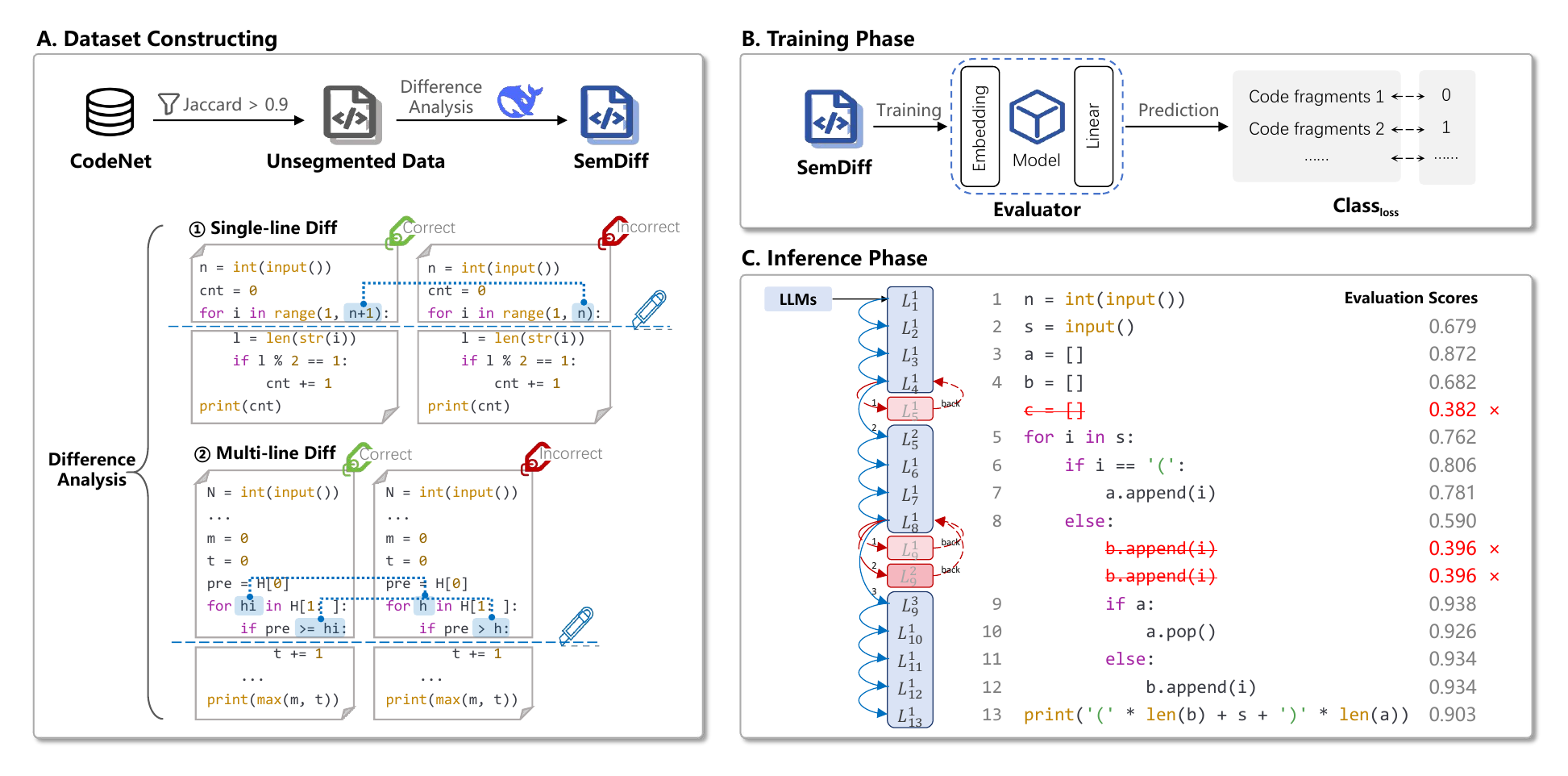}
    \caption{SemGuard Architecture, consisting of three components: (1) Dataset Construction, (2) Training Phase, (3) Inference Phase. In Phase~C, \(L_{i}^{j}\) denotes the~\(i\)-th line of code after the \(j\)-th attempt (\(j{=}1\) initial, \(j{>}1\) after backtracking).
}
    \label{fig:semGuard}
    \vspace{-4mm}
\end{figure*}

We next detail SemGuard, our framework for ensuring semantic correctness via real-time evaluation of code fragments (Figure \ref{fig:semGuard}).
The workflow comprises three stages:
\ding{182} \emph{dataset construction}, which produces line-level semantic-deviation pairs;
\ding{183} \emph{training}, where the annotated corpus is used to train a lightweight evaluator; and
\ding{184} \emph{inference}, in which the trained evaluator is embedded into the LLM’s decoding process to detect deviations on-the-fly and trigger targeted backtracking.

\subsection{Dataset Constructing}
Training a fragment-level semantic evaluator requires data that pinpoints where a program first diverges from the intended logic—information missing from existing public corpora.
We therefore build \textit{SemDiff}, a new dataset with train, validation, and test splits in which every erroneous solution is paired with a correct counterpart and tagged at the exact line of semantic deviation. These line-level annotations allow us to slice each program into \emph{prefix + faulty line + suffix} segments, providing the fine-grained supervision needed to teach the evaluator to judge the correctness of arbitrary code fragments.


As shown in Figure~\ref{fig:semGuard}, we draw our source corpus from
\textit{CodeNet}, which comprises more than 14 million submissions in
over 50 programming languages and covers roughly 4 000 competitive-
programming tasks.
We treat each submission as an ordered line sequence
\(
  C=\{\,l_{1},l_{2},\dots,l_{n}\,\},
\)
where \(l_i\) is the \(i\)-th line of code.

\textbf{Verification.}
Every solution labelled \emph{correct} in CodeNet is re-executed in a
local sandbox; samples whose output deviates from the ground-truth
answer are discarded.
For solutions labelled \emph{incorrect}, we keep only those that fail
for semantic reasons, filtering out purely syntactic or runtime faults.

\textbf{Pairing.}
For each user we pair a retained erroneous submission
\(C_{\mathrm{err}}\) with a correct one \(C_{\mathrm{corr}}\).
Let \(T_{\mathrm{corr}}\) and \(T_{\mathrm{err}}\) be their respective
\(n\)-gram sets; we compute the Jaccard similarity
\begin{equation}J\bigl(C_{\mathrm{corr}},C_{\mathrm{err}}\bigr)=
  \frac{\lvert T_{\mathrm{corr}}\cap T_{\mathrm{err}}\rvert}
       {\lvert T_{\mathrm{corr}}\cup T_{\mathrm{err}}\rvert}
\end{equation}
Pairs with \(J>0.9\) are retained, ensuring that the two versions differ
lexically as little as possible—typically by a single, localized
semantic deviation.




For every retained pair
\((C_{\mathrm{corr}},C_{\mathrm{err}})\)
we perform a line–by–line diff and record the index set
\begin{equation}
  D=\{\,i \mid l^{\mathrm{corr}}_{i}\neq l^{\mathrm{err}}_{i}\,\}
\end{equation}
whose cardinality \(\lvert D\rvert\) is the number of differing lines.
If \(\lvert D\rvert=1\) the pair differs in a single line (see \emph{Single-line Diff} case); otherwise the
changes span multiple lines (see \emph{Multi-line Diff} case).
Let \(i^{\ast}=\min D\) be the first point of deviation.
We slice each program at \(i^{\ast}\) to obtain a \emph{semantic prefix}
that will serve as a training instance:
\begin{equation}
\begin{aligned}
  S_{\mathrm{corr}} &= \langle l^{\mathrm{corr}}_{1},\,l^{\mathrm{corr}}_{2},\,\dots,\,l^{\mathrm{corr}}_{i^{\ast}} \rangle\\
  S_{\mathrm{err}}  &= \langle l^{\mathrm{err}}_{1},\,l^{\mathrm{err}}_{2},\,\dots,\,l^{\mathrm{err}}_{i^{\ast}} \rangle
\end{aligned}
\label{eq:semantic-segments}
\end{equation}
The two prefixes differ only in the final line \(l_{i^{\ast}}\), giving
the evaluator a minimal contrast between a correct and an erroneous
semantic unit.

The \emph{Multi-line Diff} example illustrates a typical pair with more than one modified line.  
Here the first mismatch is merely a renamed variable and leaves the program semantics intact, whereas the second mismatch alters the branch condition and is the true source of the semantic fault.

Because pinpointing the fault line \(i^{\ast}\) is non-trivial when several lines differ, we enlist an LLM (DeepSeek-V3 \cite{liu2024deepseek}) with strong code understanding.  
Directly feeding the raw pair to the model, however, often yields unreliable results.  
To improve precision we supply the high-similarity pair as context and \emph{explicitly instruct the model to compare the erroneous version against the correct one}.  
The prompt template is shown in Table~\ref{tab:prompt}. 
This guided comparison enables the model to highlight the single line \(i^{\ast}\) that causes the semantic deviation, which we then use as the cut point.

Combining this LLM-assisted localization for multi-line diffs with the straightforward rule for single-line diffs yields the final, line-level annotations that make up our \textit{SemDiff} corpus.


\begin{table}[ht]
\centering
\begin{tcolorbox}[
    colback=blue!3!white,
    colframe=blue!50!black,
    title=\textbf{Prompt Template for Semantic Divergence Identification},
    fonttitle=\bfseries\small,
    rounded corners,
    boxrule=1pt,
    left=3pt,
    right=3pt,
    top=2pt,
    bottom=2pt,
    width=0.99\linewidth
]

\setstretch{0.8}
\footnotesize 
\noindent Please act as a senior programmer. Based on the programming question (\textcolor{blue!70!black}{\textbf{QUESTION}}), identify the erroneous line in the code (\textcolor{red!70!black}{\textbf{RESPONSE 1}}). Refer to the correct code (\textcolor{green!70!black}{\textbf{RESPONSE 2}}) to make the judgment.

\vspace{0.3em}
\noindent It is known that (\textcolor{red!70!black}{\textbf{RESPONSE 1}}) is the incorrect code, and (\textcolor{green!70!black}{\textbf{RESPONSE 2}}) is the very similar correct code. Based on your judgment, output the line number of the initial erroneous line in (\textcolor{red!70!black}{\textbf{RESPONSE 1}}). Please do not provide any other explanations, just return the line number of the initial error.

\begin{tcolorbox}[
    colback=yellow!10!white,
    colframe=orange!60!black,
    boxrule=0.5pt,
    left=2pt,
    right=2pt,
    top=1pt,
    bottom=1pt,
    sharp corners
]
\textbf{**Note**}: 
You must deeply understand the semantic information of the code. When referencing the correct code, do not perform line-by-line comparison and directly return the line number of the first differing line.
\end{tcolorbox}


\begin{tabular}{@{}p{0.25\textwidth}p{0.8\textwidth}@{}}
\textcolor{blue!70!black}{\textbf{[QUESTION]}} & \textcolor{blue}{\texttt{\{Programming Question\}}} \\[0.3em]

\textcolor{red!70!black}{\textbf{[RESPONSE 1]}} & 
\begin{minipage}[t]{0.8\textwidth}
\texttt{[The start of \textbf{RESPONSE 1}]} \\
\textcolor{blue}{\texttt{\{The Incorrect Response\}}} \\
\texttt{[The end of \textbf{RESPONSE 1}]}
\end{minipage} 

\\[0.1em]
\textcolor{green!70!black}{\textbf{[RESPONSE 2]}} & 
\begin{minipage}[t]{0.8\textwidth}
\texttt{[The start of \textbf{RESPONSE 2}]} \\
\textcolor{blue}{\texttt{\{The Correct Response\}}} \\
\texttt{[The end of \textbf{RESPONSE 2}]}
\end{minipage} 

\\[0.1em]
\textcolor{purple!70!black}{\textbf{[OUTPUT]}} &  \\
\end{tabular}

\end{tcolorbox}
\caption{Prompt for identifying semantic deviation in code pairs using LLM.}
\label{tab:prompt}
\end{table}





\subsection{Semantic-Evaluator Training}

Phase B of Figure \ref{fig:semGuard} outlines the training pipeline.
The evaluator is a lightweight \textit{LLM, binary head}
fine-tuned to decide whether a given fragment is semantically correct.

\textbf{Backbone choice.}
Any off-the-shelf LLM can serve as the backbone.  
Because the evaluator must run inside the decoding process, we prefer models with \emph{tens or low hundreds of millions} of parameters to keep latency and memory overhead acceptable; in our experiments we adopt the 1.3B-parameter DeepSeek-Coder as a representative option.

\textbf{Architecture.}
We tokenize the sequence \( S = \langle l_{1}, \dots, l_{n} \rangle \) and the question to obtain a contextual embedding \( V \in \mathbb{R}^{n \times d} \)
from the frozen backbone.
We take the CLS / BOS token representation, pass it through a linear layer,
and apply a sigmoid to produce the correctness probability
\(p=\sigma\bigl(W V_{\text{CLS}}+b\bigr)\).

\textbf{Objective.}
The model is trained with the standard binary cross-entropy loss:
\begin{equation}
\mathcal{L}
  =-\frac{1}{k}\sum_{i=1}^{k}
    \Bigl(y_i\log p_i + (1-y_i)\log(1-p_i)\Bigr)
\end{equation}
where \(y_i\in\{0,1\}\) is the ground-truth label (“incorrect” / “correct”)
for the \(i\)-th fragment and \(p_i\) is the predicted probability.

\textbf{Remarks.}
Using a smaller LLM keeps inference cost low, while fine-tuning on
\emph{SemDiff} equips the model with precise, line-level semantic
discrimination.  In practice, we find that LLMs below 2B parameters
strike a good balance between speed and accuracy; larger backbones offer
diminishing returns once the fragment-level supervision is provided.

\subsection{Inference Phase}

Let the partially generated program be
\(L_{1:t}=\{L_{1},L_{2},\dots,L_{t}\}\),
where \(L_{t}\) is the \(t\)-th line produced so far.  
Starting with the second line, we feed each growing prefix into the
semantic evaluator, which returns a confidence score
\(s_{t}\in[0,1]\).
If \(s_{t}>0.5\) the prefix is accepted; otherwise a semantic deviation
is flagged and the system rolls back to the \emph{beginning of the
current line}, because that line is the first to turn the score from
positive to negative.

\paragraph{Token-penalty scheme.}
To avoid repeating the same error, we attenuate the probability of the
first non-indented token on the faulty line.  
Let \(p=\{p_{1},\dots,p_{n}\}\) be the original next-token distribution
and \(k\) index the token just generated.
With penalty factor \(\lambda\in(0,1)\) we set
\begin{equation}
p'_{i}=
\begin{cases}
\lambda\,p_{k}, & i=k,\\[4pt]
p_{i},           & i\ne k,
\end{cases}
\qquad
p''_{i}= \frac{p'_{i}}{\sum_{j}p'_{j}}
\end{equation}
The decoder then resamples the line up to \(N\) times.
If an attempt \(j\) achieves \(s_{t}^{(j)}>0.5\) we accept it
immediately; otherwise we keep the trial with the highest score:
\begin{equation}
L^{\star}_{t}=\arg\max_{j\in\{1,\dots,N\}}s_{t}^{(j)}
\end{equation}
The prefix is updated to
\(L_{1:t}\!\leftarrow\!\{L_{1},\dots,L^{\star}_{t}\}\) and generation
continues with line \(t{+}1\).

\paragraph{Illustrative trace.}
Stage~C of Figure~\ref{fig:semGuard} shows a run on \textit{SemDiff}.  
When the first generation  of line~5 (\texttt{C = []}) is appended, the prefix
score drops to 0.38, triggering a rollback.  Penalizing the token
\texttt{c} and resampling yields \texttt{for i in s:} with
\(s_{5}=0.76\), so generation proceeds.  
At line~9 the same procedure is invoked twice: each time the token
\texttt{b} is penalized, its probability wanes, and the third resample
finally restores a score above the threshold, after which the program
completes without further deviations.

\section{Experimental Design}

\subsection{Research Questions}


To validate the effectiveness of SemGuard, we propose the following six research questions (RQs): 

\setlist[itemize]{left=0pt}
\begin{itemize}

\item\textbf{RQ1: Compared to Baseline Approaches.} How does SemGuard perform compared to baseline approaches in code generation?



\item\textbf{RQ2: Performance Across Different LLMs.} How does SemGuard perform across different LLMs?



\item\textbf{RQ3: Transferability.} How does SemGuard's transferability?



\item\textbf{RQ4: Performance on Other Programming Language.} How does SemGuard perform on other programming language?



\item\textbf{RQ5: Ablation Study.} How does each component of SemGuard contribute to its effectiveness?



\item\textbf{RQ6: Cost and Efficiency of SemGuard.} How does SemGuard perform in terms of cost and efficiency during code generation?

\end{itemize}

\subsection{Datasets and Evaluation Metric}

In this paper, we conduct experiments on four datasets, including two custom-constructed datasets and two widely used benchmarks in code generation. The custom datasets are designed to support the training and evaluation of partial code semantic evaluators, while the public datasets help assess the generalization of our method.

\setlist[itemize]{left=0pt}
\begin{itemize}


\item \textbf{\textit{SemDiff}} contains \textbf{998} competition-level CodeNet problems and a total of \textbf{123,522} annotated code fragments, split into \textbf{114,098(437)} training, \textbf{5,784(441)} validation, and \textbf{3,640(120)} test samples.  The corpus is tailored for training partial-code semantic evaluators and evaluating their ability to judge the correctness of incomplete programs.


\item \textbf{\textit{SemDiff-Java}} is the Java counterpart of \textit{SemDiff}.  
It comprises \textbf{99,882} training, \textbf{4,262} validation, and \textbf{3,672} test samples.  
We use this corpus to train a Java-specific partial-code evaluator and to assess SemGuard’s cross-language effectiveness.


\item \textbf{\textit{MBPP}}~\cite{austin2021programsynthesis} comprises \textbf{974} Python tasks  
(\textbf{374} train, \textbf{90} validation, \textbf{500} test, and \textbf{10} few-shot samples).  
We employ the \textbf{500} test tasks to gauge SemGuard’s transferability.


\item \textbf{\textit{LiveCodeBench}}~\cite{jain2024livecodebench} continuously harvests real-world problems from LeetCode, AtCoder, and CodeForces, time-stamping each task to avoid train–test leakage.  
For an uncontaminated evaluation set, we use only tasks collected between \textbf{1 July 2024} and \textbf{1 April 2025}.
\end{itemize}


Following previous research, we use the Pass@$k$~\cite{chen2021evaluating} metric to measure the functional correctness of code generated by LLMs. For each task, $n \geq k$ code samples are generated, and the number of samples passing the test cases is computed $c \leq n$. The Pass@$k$ is then estimated using the formula:

\begin{equation}
\text { Pass@$k$ }=\underset{\text { Problems }}{\mathrm{E}}\left[1-\frac{\binom{n-c}{k}}{\binom{n}{k}}\right]
\end{equation}

In our experiments, we report only Pass@1, as it provides the most straightforward and informative measure of the LLMs' ability to generate correct code in a single attempt. To enhance the robustness and credibility of our evaluation, we run each experiment three times and use the average result.

\subsection{Base LLMs and Baselines}





We conduct experiments using several open-source LLMs, including DeepSeek-Coder (1.3B\&6.7B)~\cite{guo2024deepseek}, CodeLlama (7B)~\cite{roziere2023code}, QwenCoder (3B\&7B)~\cite{hui2024qwen2}, and StarCoder2 (3B\&7B)~\cite{li2023starcoder}, all of which have been widely used in code generation tasks. To assess the performance of our approach, We compare it with three baseline methods that intervene in the decoding or inference stage of LLMs, among which ROCODE is the current state-of-the-art (SOTA) method.

\setlist[itemize]{left=0pt}
\begin{itemize}
\item \textbf{Temperature Sampling}~\cite{caccialanguage} has become widely used, where a temperature coefficient T is used to control the randomness of sampling. A higher temperature T leads to more randomness in token selection, resulting in greater diversity, while a lower temperature T makes token selection more specific, yielding more deterministic results.

\begin{equation}
P^{\prime}(w)=\frac{\exp \left(\log \left(P\left(w \mid w_{<t}\right)\right) / T\right)}{\sum_{w^{\prime}} \exp \left(\log \left(P\left(w^{\prime} \mid w_{<t}\right)\right) / T\right)}
\end{equation}

\item \textbf{Sampling + Filtering}~\cite{chen2023codet} is a code generation method that combines sampling and filtering. First, an LLM generates a large number of code samples, covering various possible code implementations. Then, these code samples are rigorously filtered by executing test cases, selecting only the code that correctly executes and passes the tests.


\item \textbf{ROCODE}~\cite{jiang2024rocode} is a constrained decoding baseline that combines program execution, lightweight static analysis, and iterative backtracking.  
      After each decoding attempt, the generated program is run against public test cases; failing traces are analysed to locate the first suspicious line.  
      The decoder then rolls back to that line and regenerates the remaining code while applying an \emph{exponentially decaying token-penalty} to discourage repetition of the same faulty path.  
      This backtrack-and-regenerate cycle continues until all tests pass or a preset step limit is reached.

\end{itemize}

\subsection{Training and Inference Settings}


\mypara{Generation Model}
During experimentation we sample from the base model with a temperature of 0.8 and a Top-$p$ of 0.95. We observe that, under these settings, the model’s performance metrics fluctuate markedly and remain relatively low. Because our benchmarks comprise competition-level tasks, we wish to validate the proposed method under a stronger baseline that exhibits fewer compilation and runtime failures. To stabilize the metrics at a higher level, we select the Top-20 ranked solutions in the \textit{SemDiff} training split (8,740 samples in total) and fine-tune the base generator for five epochs with a learning rate of 2e-5 with LoRA. The LoRA configuration is $r = 8$, $\alpha = 32$, $\text{dropout} = 0.1$, with adapters applied to \texttt{q\_proj} and \texttt{v\_proj}.
We then choose the checkpoints that yield the most stable results. All methods—including SemGuard and all baselines—share identical base generators and the same LoRA-fine-tuned checkpoints, ensuring fairness. We run three independent trials with each method and report the average to mitigate sampling variance. In the experiment, the penalty factor is 0.8, with a maximum of 3 samplings.

\mypara{Semantic Evaluator Model}We choose the CodeT5 encoder (only used in RQ 5) and DeepSeekCoder-1.3B as the base architectures for training semantic evaluators at different parameter scales. Because the training corpus contains incomplete programs with subtle semantic differences, we fine-tune each base model for 15 epochs using a learning rate of 6e-5 and a batch size of 50 to obtain a more discriminative evaluator. All experiments run on four RTX A6000 (48 GB) GPUs.


\section{Experimental Results and Analysis}
\subsection{Compared to Baseline Approaches.}


\textit{\textbf{Setup.}}
On the \textit{SemDiff} test set we compare five decoding strategies—Temperature Sampling (Temp.), Sampling + Filtering (S + F), ROCODE, and our two variants, SemGuard-Random (SG-R) and SemGuard-Penalty (SG-P)—using DeepSeekCoder-6.7B and QwenCoder-7B as the underlying LLMs.
SemGuard-Random backtracks to the faulty line but applies no token-level penalty, whereas SemGuard-Penalty adds a focused penalty to the first non-indented token of that line; the three baselines follow their original implementations.

\textit{\textbf{Results and Analyses.}} The experimental results, summarized in Table~\ref{tab:table1}, SemGuard-Penalty outperforms all baseline methods, including the current SOTA baseline method, ROCODE. Furthermore, SemGuard-Random, which relies on random decision-making during backtracking, achieves only modest performance improvements. This observation further underscores the effectiveness of the penalty strategy employed in SemGuard-Penalty for enhancing code generation quality.




















\begin{table}[t]
  \centering
  \caption{Pass@1 (\%) on \textit{SemDiff}.  Best scores in \textbf{bold}, second best \underline{underlined}.}
  \label{tab:table1}
  \renewcommand{\arraystretch}{1.25}
  \small
  \setlength{\tabcolsep}{3pt}  
  \begin{tabular}{lcc}
    \toprule
    \textbf{Method} & \textbf{DeepSeekCoder-6.7B} & \textbf{QwenCoder-7B} \\
    \midrule
    Temperature Sampling    & 30.28 & 30.83 \\
    Sampling + Filtering    & 33.33 & 34.17 \\
    ROCODE                  & \underline{35.83} & \underline{37.50} \\
    \midrule
    SemGuard-Random         & 33.33 & 34.16 \\
    SemGuard-Penalty & \textbf{38.06} & \textbf{38.34} \\
    \bottomrule
  \end{tabular}
\end{table}

\begin{figure}[!t]
    \centering
    \includegraphics[width=\linewidth]{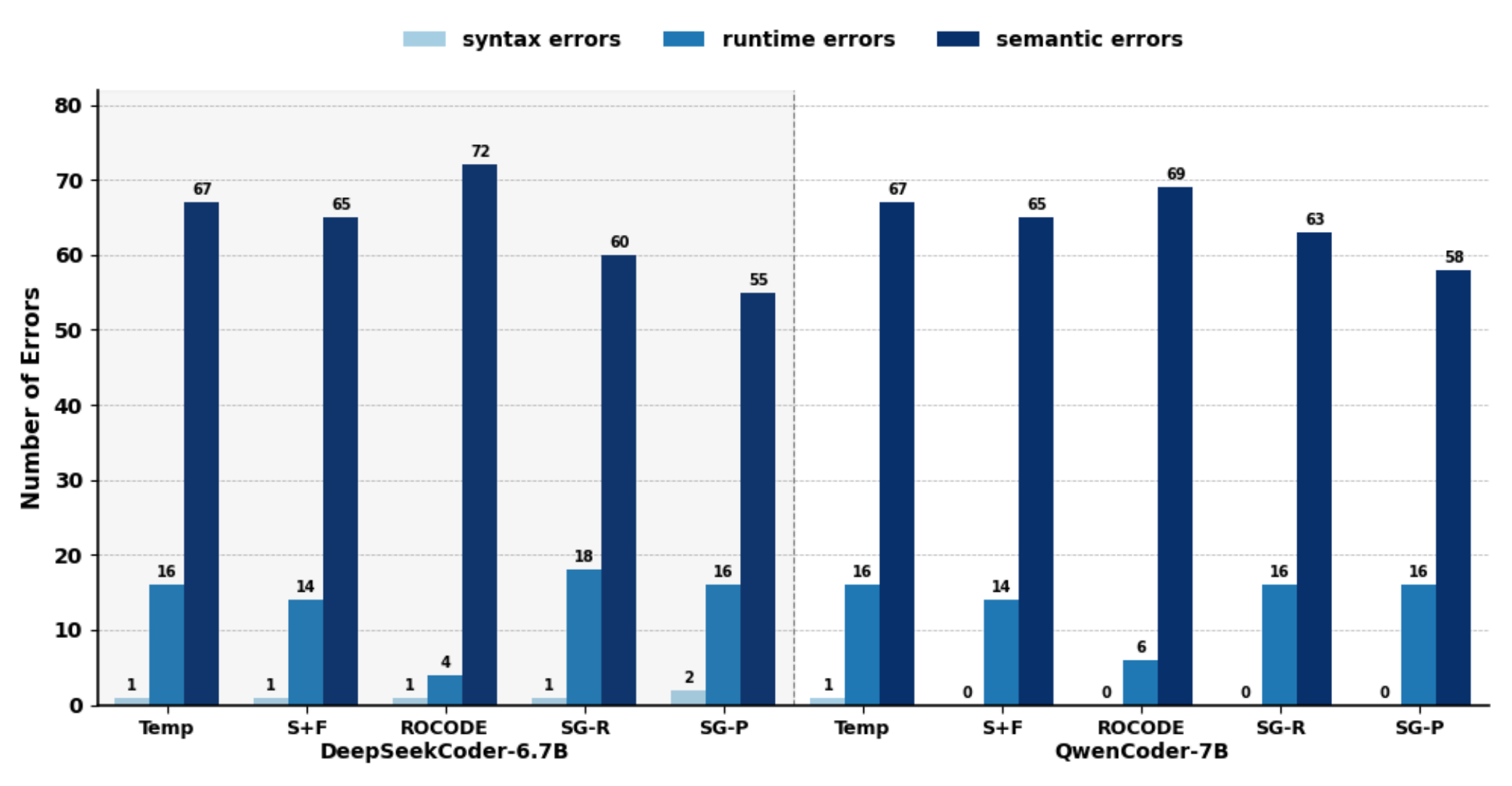}
    \caption{Counts of syntax, runtime, and semantic errors under different decoding strategies.}
    \label{fig:last}
\end{figure}

Figure~\ref{fig:last} shows the number of problems involving three types of errors—syntax errors, runtime errors, and semantic errors—produced by SemGuard and the baseline methods. SemGuard-Random produces fewer semantic errors than all baseline models, while SemGuard-Penalty yields the fewest semantic errors overall. When compared to ROCODE, which shows strong performance in reducing compilation and runtime errors, SemGuard-Penalty demonstrates a clear advantage in mitigating semantic errors. Given the relatively low incidence of compilation and runtime errors—especially in larger models—efforts should increasingly focus on reducing semantic inaccuracies. Notably, even in the QwenCoder-7B setting, SemGuard-Penalty performs comparably to ROCODE in terms of Pass@1, but with significantly fewer semantic errors. This result further highlights the advantage of our method in enhancing the semantic correctness of generated code, validating its practical potential in real-world tasks.

\begin{tcolorbox}[size=title,breakable]
\textit{\textbf{Answer to RQ1:} \textcolor{black}{Compared to baseline methods ROCODE, SemGuard excels in reducing semantic errors, significantly lowering their occurrence, while eliminating the need for test cases to execute the code.}}
\end{tcolorbox}

\subsection{RQ2: Performance Across Different LLMs.}

\begin{table}[t]
  \centering
  \caption{Pass@1 (\%) on \textit{SemDiff}.  Best in \textbf{bold}, second best \underline{underlined}.}
  \label{tab:table_2}
  \renewcommand{\arraystretch}{1.25}
  \resizebox{\linewidth}{!}{%
  \begin{tabular}{lcccc}
    \toprule
    \textbf{Model} & \textbf{Temp.} & \textbf{ROCODE} & \textbf{SG-R} & \textbf{SG-P} \\
    \midrule
    DeepSeekCoder-6.7B & 30.28 & \underline{35.83} & 33.33 & \textbf{38.06} \\
    QwenCoder-3B       & 22.22 & 23.33 & \underline{23.34} & \textbf{26.11} \\
    QwenCoder-7B       & 30.83 & \underline{37.50} & 34.16 & \textbf{38.34} \\
    StarCoder2-3B      & 16.11 & \underline{18.33} & 17.50 & \textbf{19.44} \\
    StarCoder2-7B      & 21.11 & \underline{23.05} & 22.78 & \textbf{25.83} \\
    CodeLlama-7B       & 15.28 & \underline{17.77} & 15.78 & \textbf{18.05} \\
    \bottomrule
  \end{tabular}}
\end{table}


\textit{\textbf{Setup.}} 
We benchmark four decoding strategies—SemGuard-Random, SemGuard-Penalty, Temperature Sampling, and ROCODE—on the \textit{SemDiff} test set using six open-source LLM checkpoints:  
DeepSeekCoder-6.7B; QwenCoder-3B/7B; StarCoder2-3B/7B; and CodeLlama-7B.  
All runs share the same generation hyper-parameters as in RQ1.


\textit{\textbf{Results and Analyses.}}  
Table~\ref{tab:table_2} reports the empirical results.  
Across every backbone and parameter scale, SemGuard-Penalty consistently surpasses both ROCODE and conventional Temperature Sampling, confirming the robustness of our constrained decoding design over a wide spectrum of model families.  
It likewise exceeds SemGuard-Random on all checkpoints, underscoring the necessity of the token-penalty mechanism.  

For a finer comparison we include QwenCoder and StarCoder models at 3B and 7B scales.  
The larger 7B variants yield markedly larger absolute gains, indicating that SemGuard benefits from the stronger priors of higher-capacity models and thereby suppresses semantic errors more effectively.  
Notably, the DeepSeekCoder model improved by 25.69\%, and the QwenCoder model improved by 24.36\%, representing the largest performance gains among all the models compared. Moreover, these models also have the highest original metrics. This phenomenon further suggests that our method is particularly effective in reducing semantic errors when starting with a model with higher initial capability, thereby achieving more significant performance improvements.

\begin{tcolorbox}[size=title,breakable]
\textit{\textbf{Answer to RQ2:} \textcolor{black}{SemGuard performs exceptionally well across models of different sizes. The higher the accuracy of the semantic evaluator and the more powerful the base generation model, the stronger SemGuard's ability to reduce semantic errors in code generation.}}
\end{tcolorbox}

\subsection{RQ3: Transferability.}


\textit{\textbf{Setup.}}  
We evaluate SemGuard’s transferability on two unseen benchmarks—\textit{MBPP} and \textit{LiveCodeBench}—which we deliberately exclude when training the semantic evaluators.  
Both SemGuard variants (SemGuard-Random and SemGuard-Penalty) are compared with Temperature Sampling and ROCODE across four 7B-scale LLMs.
\textit{\textbf{Results and Analyses.}}  
Table~\ref{tab:table-generalization} shows that SemGuard-Penalty achieves the best Pass@1 on seven of the eight (\emph{model}, \emph{benchmark}) combinations, outperforming Temperature Sampling, ROCODE, and SemGuard-Random across both \textit{MBPP} and \textit{LiveCodeBench}.  
The sole exception occurs on \textit{LiveCodeBench} with CodeLlama-7B, where ROCODE attains a slightly higher score (8.73 vs 8.28).  
On \textit{MBPP}, whose tasks are generally short and structurally simple, the gap between SemGuard-Penalty and SemGuard-Random is modest (\(\approx 0.5\text{--}1.0\) points), yet SemGuard-Penalty still surpasses ROCODE for every backbone, indicating that our token-penalty strategy remains beneficial even when semantic deviations are shallow.

In contrast, \textit{LiveCodeBench} features longer, algorithm-intensive problems.  Here ROCODE gains more from its test-case oracle, particularly on CodeLlama-7B, but SemGuard-Penalty still delivers the top result for DeepSeekCoder-6.7B, QwenCoder-7B, and StarCoder2-7B.  This pattern suggests that evaluator-guided rollback scales favourably with model capacity and retains high efficacy without relying on external test cases.  
Overall, the results confirm SemGuard’s strong transferability across unseen datasets and diverse LLM families, while highlighting that high-quality public tests can occasionally narrow—or, for one setting, invert—the margin between SemGuard and ROCODE.

\begin{table}[t]
  \centering
  \caption{Pass@1 (\%) on unseen benchmarks.  Best in \textbf{bold}, second best \underline{underlined}.}
  \label{tab:table-generalization}
  \renewcommand{\arraystretch}{1.25}
  \resizebox{\columnwidth}{!}{%
  \begin{tabular}{lcccc}
    \toprule
    \textbf{Model} & \textbf{Temp.} & \textbf{ROCODE} & \textbf{SG-R} & \textbf{SG-P} \\
    \midrule
    \multicolumn{5}{c}{\textit{MBPP}} \\
    \cmidrule(lr){1-5}
    DeepSeekCoder-6.7B & 53.87 & 56.53 & \underline{57.20} & \textbf{58.20} \\
    QwenCoder-7B       & 61.60 & 63.53 & \underline{63.60} & \textbf{64.20} \\
    StarCoder2-7B      & 44.27 & 47.53 & \underline{48.73} & \textbf{49.20} \\
    CodeLlama-7B       & 38.73 & \underline{42.80} & 42.73 & \textbf{43.20} \\
    \midrule
    \multicolumn{5}{c}{\textit{LiveCodeBench}} \\
    \cmidrule(lr){1-5}
    DeepSeekCoder-6.7B & 7.68  & \underline{9.06} & 8.75  & \textbf{10.04} \\
    QwenCoder-7B       & 8.62  & 9.21  & \underline{9.57}  & \textbf{10.87} \\
    StarCoder2-7B      & 6.74  & \underline{7.80} & 7.20  & \textbf{8.74} \\
    CodeLlama-7B       & 5.56  & \textbf{8.73} & 7.09  & \underline{8.28} \\
    \bottomrule
  \end{tabular}}
\end{table}


\begin{tcolorbox}[size=title,breakable]
\textit{\textbf{Answer to RQ3:} \textcolor{black}{SemGuard generalizes well across unseen benchmarks.  
On the simpler \textit{MBPP} tasks, its penalty mechanism yields modest but consistent gains;  
on the more demanding \textit{LiveCodeBench} problems, the same mechanism delivers the largest improvements among all methods, confirming our robustness to task complexity.}}
\end{tcolorbox}

\subsection{RQ4: Performance on Other Programming Language.}


\textit{\textbf{Setup.}}  
Beyond Python, we evaluate SemGuard on Java.  Following the same pipeline as \textit{SemDiff}, we construct \textit{SemDiff-Java}; the only departure is that we adopt the ``comparative” split directly, because Java submissions are longer and exhibit extensive near-duplicates, making LLM–assisted splitting prohibitively expensive. We compare SemGuard-Random, SemGuard-Penalty, and Temperature Sampling on \textit{SemDiff-Java} with four 7B-scale backbones—DeepSeek-Coder-6.7B, QwenCoder-7B, StarCoder2-7B, and CodeLlama-7B.  
We omit ROCODE: it does not discuss the applicability to the Java language and does not disclose the implementation details of C++, reproducing its method would require significant engineering effort and may introduce confounding biases.

\begin{table}[t]
  \centering
  \caption{Pass@1 (\%) on \textit{SemDiff-Java}.}
  \label{tab:table-java}
  \renewcommand{\arraystretch}{1.2}
  \normalsize  
  \begin{tabular}{lccc}
    \toprule
    \textbf{Model} & \textbf{Temp.} & \textbf{SG-R} & \textbf{SG-P} \\
    \midrule
    DeepSeekCoder-6.7B & 33.58 & 36.32 & \textbf{42.53} \\
    QwenCoder-7B       & 33.94 & 37.39 & \textbf{40.94} \\
    StarCoder2-7B      & 30.30 & 31.71 & \textbf{34.90} \\
    CodeLlama-7B       & 22.08 & 23.95 & \textbf{26.43} \\
    \bottomrule
  \end{tabular}
\end{table}

\textit{\textbf{Results and Analyses.}}  
Table~\ref{tab:table-java} confirms that SemGuard transfers well to Java.  
Across all four backbones, SemGuard-Random increases Pass@1 by roughly two to three points over Temperature Sampling, showing that evaluator-guided backtracking is valuable even without token penalties.  Adding the penalty mechanism delivers an additional 3-6 points boost, so SemGuard-Penalty achieves 15–27 \% relative gains overall.

A closer look at the per-model numbers reveals two consistent trends.  \textbf{First}, DeepSeekCoder-6.7B and QwenCoder-7B realize the largest absolute lifts, echoing their strong Python-side gains in RQ2 and suggesting that higher-capacity models benefit most from fine-grained semantic feedback.  
\textbf{Second}, CodeLlama-7B still gains almost four points despite the lowest baseline (22.08 → 26.43), indicating that SemGuard remains helpful when the decoder struggles with Java’s longer, boiler-plate-heavy solutions.  
Taken together, the evidence reinforces SemGuard’s language-agnostic design: evaluator-driven rollback consistently improves Java generation without any language-specific tuning, and the token-penalty strategy scales favourably with both model capacity and task complexity.


\begin{tcolorbox}[size=title,breakable]
\textit{\textbf{Answer to RQ4:} \textcolor{black}{SemGuard generalizes beyond Python, delivering consistent Pass@1 improvements on Java and thus confirming its language-agnostic adaptability.}}
\end{tcolorbox}

\subsection{RQ5: Ablation Study.}


\textit{\textbf{Setup.}}
We factorize SemGuard into two dimensions.
For \emph{semantic evaluator capacity} we use a small CodeT5-770M evaluator and a larger DeepSeekCoder-1.3B evaluator.
For the \emph{backtracking policy} we test four variants: Full-Restart Backtracking, which rolls back to the beginning of the file and regenerates from scratch; Exponentially-Decaying Penalty (EDP) from ROCODE, which retains the prefix but imposes a linearly shrinking penalty toward the rollback point; SemGuard-Random and SemGuard-Penalty.
Combining the two evaluators with the four policies yields six configurations, enabling us to disentangle the effects of evaluator scale and penalty design on SemGuard’s performance.


\textit{\textbf{Results and Analyses.}}  
Table~\ref{tab:table6} shows a clear separation between evaluator capacity and back-tracking policy.  
Using the smaller CodeT5-770M evaluator depresses Pass@1 regardless of backtracking choice, indicating that a 770M model lacks the representational power to judge subtle semantic deviations. By contrast, the stronger DeepSeek-1.3B evaluator raises the ceiling for every policy and, when paired with our token-penalty scheme (SemGuard-Penalty), achieves the best result, surpassing Temperature Sampling by nearly eight points.  

The comparison among backtracking policies is equally telling.  SemGuard-Penalty outperforms EDP even though both reuse the same evaluator, suggesting that a single, line-focused penalty preserves more correct context than EDP’s coarse, file-wide decay.  SemGuard-Random and Full Restart BackTracking deliver only modest gains: the former lacks directional guidance, while the latter discards useful prefixes and thus wastes computation. Together these observations confirm that \ding{182} evaluator precision is crucial for reliable semantic control and \ding{183} a targeted, line-level penalty is the most effective way to exploit that precision during decoding.

\begin{table}[t]
  \centering
  \caption{
  Ablation on Evaluator Capacity \& Backtracking Policy (Generator = DeepSeek-Coder-6.7B)
  }
  \label{tab:table6}
  \renewcommand{\arraystretch}{1.25}
  \resizebox{\columnwidth}{!}{%
  \begin{tabular}{lcc}
    \toprule
    \textbf{Evaluator} & \textbf{Backtracking Policy} & \textbf{Pass@1} \\
    \midrule
    \multirow{2}{*}{CodeT5-770M}
      & SemGuard-Random & 27.50 \\
      & SemGuard-Penalty  & 28.33 \\
    \midrule
    \multirow{4}{*}{DeepSeek-1.3B}
      & Full-Restart Backtracking               & 32.97 \\
      & Exponentially-Decaying Penalty           &  35.00 \\
      & SemGuard-Random            & 33.33 \\
      & SemGuard-Penalty    &  \textbf{38.06} \\
    \midrule
    \multicolumn{2}{l}{Temperature Sampling (baseline)} & 30.28 \\
    \bottomrule
  \end{tabular}}
\end{table}


\begin{tcolorbox}[size=title,breakable]
\textit{\textbf{Answer to RQ5:} \textcolor{black}{Ablation shows that evaluator quality dominates overall performance, while the line-targeted penalty further amplifies the gain.  Accurate semantic evaluators are therefore essential for reliably steering LLMs toward correct code.}}
\end{tcolorbox}

\subsection{RQ6: Cost and Efficiency of SemGuard.}


\textit{\textbf{Setup.}}  
We quantify overhead in two dimensions: \emph{token cost}, measured as the total number of generated tokens, and \emph{latency}, measured as wall-clock time per task.  
All experiments run on the \textit{SemDiff} benchmark with DeepSeek-Coder-6.7B as the generator.  
We compare five decoding strategies—Temperature Sampling, Sampling + Filtering, ROCODE, and two SemGuard configurations—under identical hardware and hyper-parameter settings.

\begin{figure}[t]
    \centering
  \includegraphics[width=\linewidth]{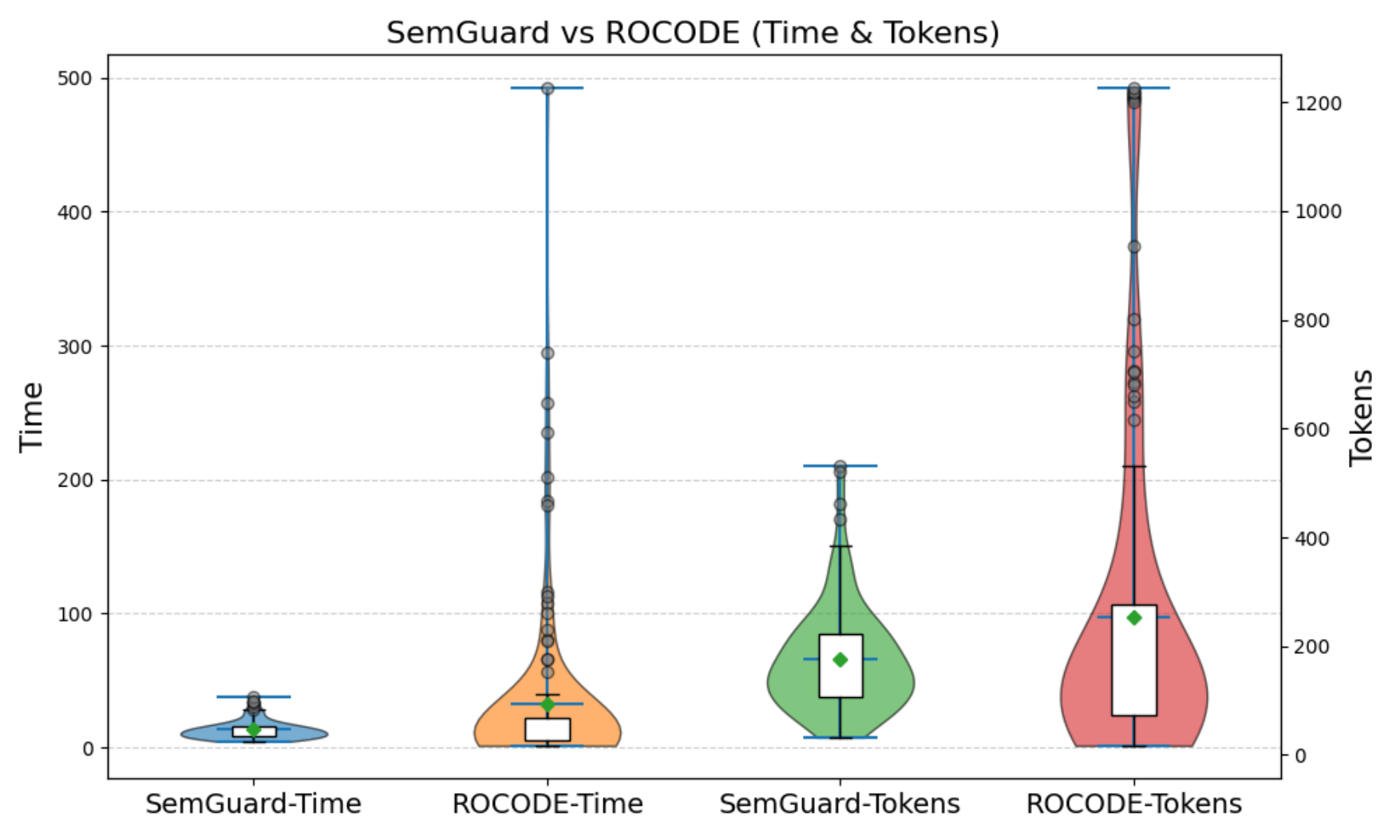}
    \caption{Time and Token Efficiency: SemGuard-Penalty vs. ROCODE.}
    \label{fig:time}
\end{figure}

\textit{\textbf{Results and Analyses.}}  
Table~\ref{tab:SemGuard-efficiency} contrasts accuracy, token cost, and latency for all methods.  
SemGuard-Penalty attains the highest Pass@1 (38.06\%), surpassing the nearest competitor ROCODE by 2.2 points while using \textbf{31\% fewer tokens} (175.6 vs 253.8) and cutting inference time by \textbf{60\%} (12.98 s vs 32.50 s).  
Temperature Sampling is the cheapest option (110.4 tokens; 6.12 s) but trails SemGuard-Penalty by nearly eight accuracy points, illustrating the cost-accuracy trade-off.  
Sampling + Filtering improves accuracy over Temperature Sampling but still lags SemGuard-Penalty and incurs an extra 55 tokens per query.  
SemGuard-Random shows that evaluator-guided backtracking alone raises Pass@1 to ROCODE’s level while halving ROCODE’s latency; adding the targeted token penalty (SemGuard-Penalty) delivers the best balance of quality and efficiency. 
Figure~\ref{fig:time} shows that SemGuard-Penalty is more efficient and stable than ROCODE, with most runs completing within 20 seconds and under 600 tokens, while a notable portion of ROCODE runs exceed one minute or 600 tokens.

Overall, SemGuard-Penalty achieves the strongest accuracy with only modest overhead—about 1.6 × the token cost and 2.1 × the latency of the minimal Temperature baseline—demonstrating that evaluator-driven rollback yields significant gains without the heavy runtime burden of test-based methods such as ROCODE.

\begin{table}[t]
  \centering
  \caption{Cost–efficiency comparison on \textit{SemDiff} (Generator = DeepSeek-Coder-6.7B).}
  \label{tab:SemGuard-efficiency}
  \renewcommand{\arraystretch}{1.25}
  \resizebox{\columnwidth}{!}{%
  \begin{tabular}{lccc}
    \toprule
    \textbf{Method} & \textbf{Pass@1 (\%)} & \textbf{Tokens} & \textbf{Time (s)} \\
    \midrule
    Temperature Sampling     & 30.28 & \textbf{110.4} & \textbf{6.12} \\
    Sampling + Filtering     & 33.33 & 230.6          &  8.38 \\
    ROCODE                   & 35.83 & 253.8          & 32.50 \\
    \midrule
    SemGuard-Random          & 33.33 & 172.6          & 13.44 \\
    SemGuard-Penalty  & \textbf{38.06} & 175.6 & 12.98 \\
    \bottomrule
  \end{tabular}}
\end{table}


\begin{tcolorbox}[size=title,breakable]
\textit{\textbf{Answer to RQ6:} \textcolor{black}{SemGuard achieves top Pass@1 at modest cost—over 50\% less runtime than ROCODE and near-baseline token use—offering a favourable accuracy–cost trade-off.}}
\end{tcolorbox}

\section{Discussion}

\subsection{False-Positive Rate of Partial-Code Judgments}
We analyze the false-positive rate (FPR) of partial-code judgments because unnecessary rejections translate into extra backtracking, tokens, and latency, thereby determining practical usability.
We sample 30 tasks per dataset (SemDiff, MBPP, LiveCodeBench) with DeepSeekCoder-7B. For each method (SemGuard-Penalty, ROCODE), we collect all \emph{rollback events} during decoding, i.e., rejected completions. For a task with $N$ such events, let $M$ be the number of flagged partial-code segments that are in fact acceptable; the per-task false-positive rate is $\mathrm{FPR}=M/N$. To reduce adjudication cost, we use a hybrid protocol: DeepSeekCoder-7B first attempts to complete each flagged partial segment (temperature 0.8; 100 candidates); if any completion passes the available tests, the segment is labeled acceptable (i.e., a false positive). Otherwise, the sample proceeds to manual review by three graduate annotators ($\ge$3 years Python/Java and familiarity with competitive-programming tasks) who judge independently and resolve disagreements by adjudication.



\begin{figure}[t]
  \centering

  \begin{subfigure}{\linewidth}
    \centering
    \includegraphics[width=\linewidth]{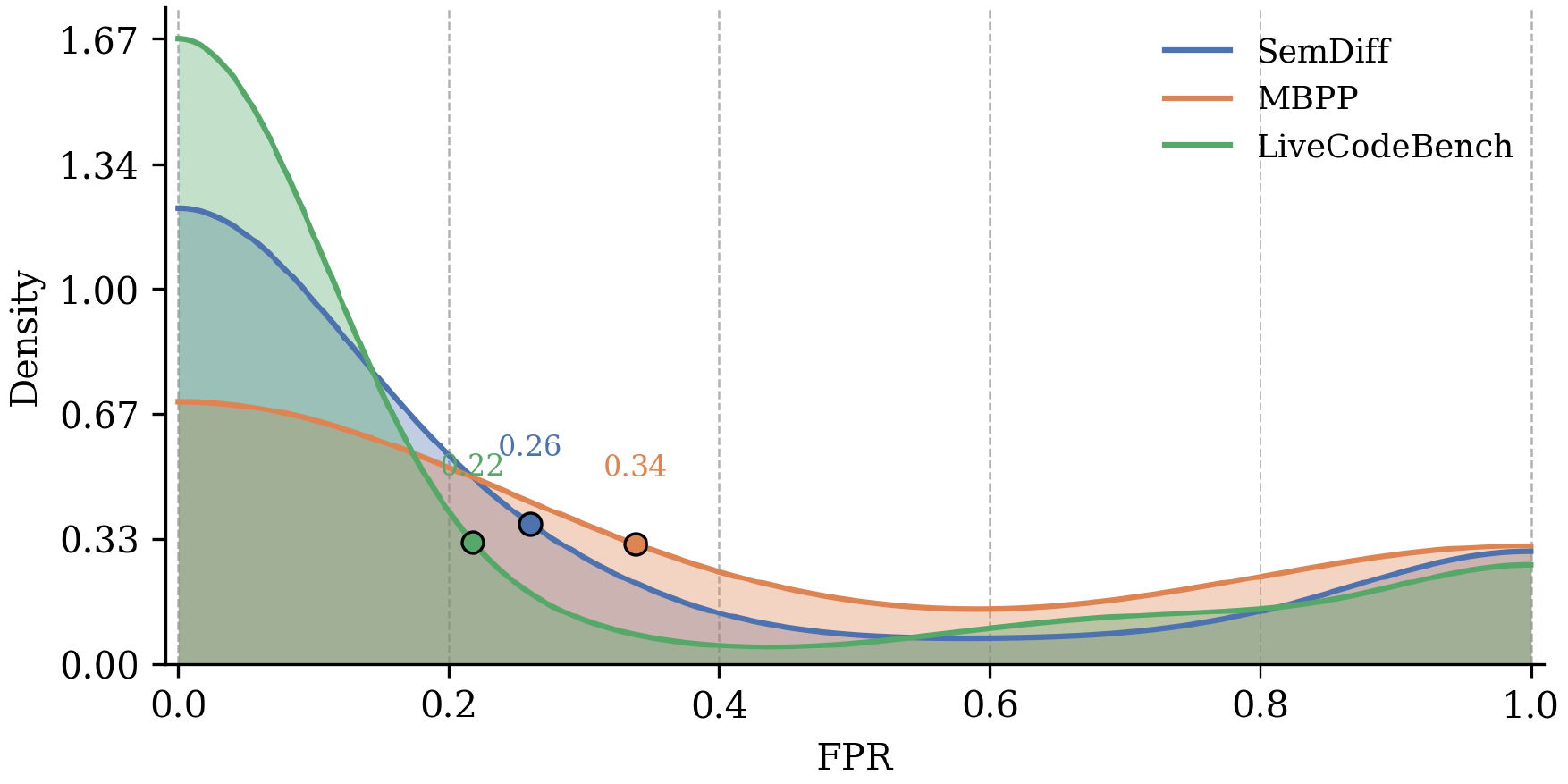}
    \caption{Per-task FPR density for SemGuard-Penalty.}
    \label{fig:kde_semguard}
  \end{subfigure}

  \vspace{0.6em} 

  \begin{subfigure}{\linewidth}
    \centering
    \includegraphics[width=\linewidth]{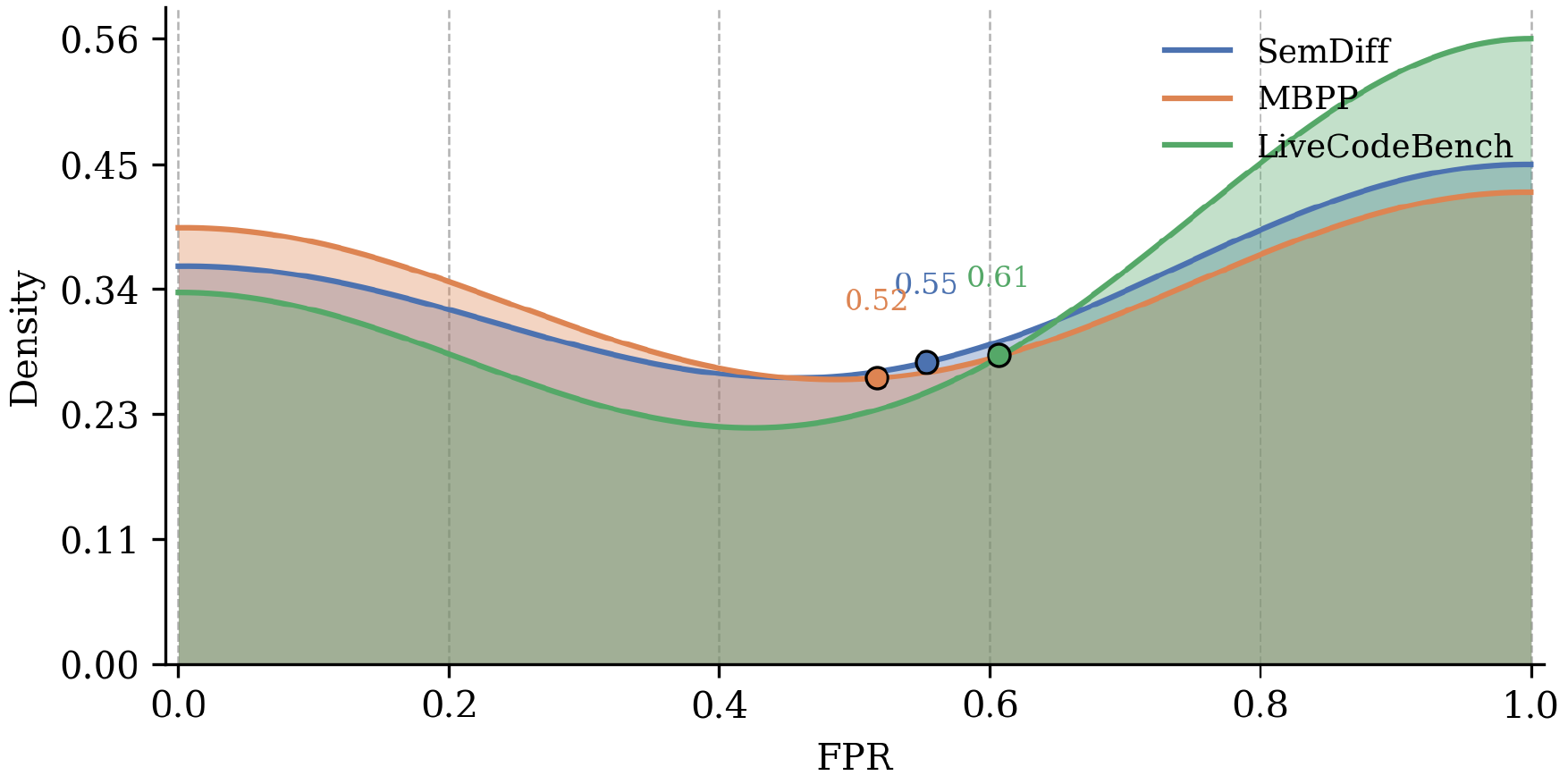}
    \caption{Per-task FPR density for ROCODE.}
    \label{fig:kde_rocode}
  \end{subfigure}

  \caption{FPR Density.}
  \label{fig:kde_two}
\end{figure}


Across datasets (Figure~\ref{fig:kde_semguard}), SemGuard-Penalty concentrates in the low--false-positive region: most mass lies at $\leq 0.40$, with rare extreme highs. MBPP shifts rightward relative to SemDiff and LiveCodeBench, likely because its snippets are very short (often just 2--3 lines), offering limited local context. In contrast (Figure~\ref{fig:kde_rocode}), ROCODE shows a mean false positive rate above $0.50$ on all three datasets, highlighting the limits of entropy-based localization.

\textit{Link to latency.} Figure~\ref{fig:time} reports time and tokens on the \emph{full} SemDiff test set, whereas the false-positive analysis uses 30-task samples from SemDiff/MBPP/LiveCodeBench. On SemDiff, the sampled FPR for SemGuard-Penalty is consistently lower than for ROCODE. Under our decoding policy each false positive triggers a rollback and re-generation, so lower FPR implies less work. This mechanism-level expectation aligns with Figure~\ref{fig:time}: SemGuard-Penalty concentrates below 20\,s and 600 tokens, while ROCODE exhibits long tails. We therefore conclude that FPR differences are an important but not exclusive factor in the latency gap. ROCODE’s latency mainly stems from deferred semantic checking, as validation occurs only after full program generation and test execution, causing each rollback to waste many tokens.

\subsection{Scope \& Limitations}



\begin{figure}[t]
  \centering
  \includegraphics[width=\linewidth]{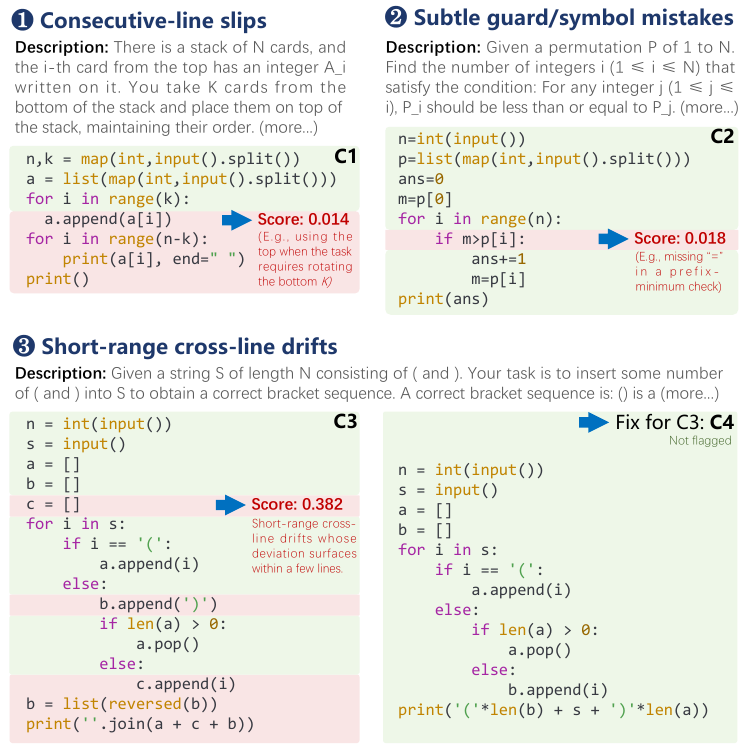} 
  \caption{SemGuard scope examples (Code C1–C4). Red: erroneous lines; Green: correct lines. Problems are taken from LiveCodeBench (\#abc368\_a) and SemDiff (\#0031, \#0104).}
  \label{fig:scope-cases}
\end{figure}

What SemGuard reliably catches. Under a prefix assumption with short local context, SemGuard is effective at: (i) \textbf{consecutive-line slips}—e.g., in Figure~\ref{fig:scope-cases} \textbf{C1}, the code rotates the top segment when the task requires rotating the bottom-$K$. This error is common in models due to the limitations of autoregressive decoding. Once an error happens, the model cannot fix it and continues generating based on that mistake; and (ii) \textbf{subtle guard/symbol mistakes}—e.g., \textbf{C2} should enforce that \(P_i\) is the minimum over the prefix \(P_{1:i}\), but line 6 is missing the $=$; and (iii) \textbf{short-range cross-line drifts} (often across non-consecutive lines)—e.g., \textbf{C3} introduces an unnecessary list \(c\), conflating roles in parenthesis balancing, whereas \textbf{C4} is the minimal fix that restores the two-list invariant (\(a\) for unmatched ``('' and \(b\) for extra ``)''), which SemGuard does not flag.

What SemGuard may miss or misjudge. Non-local logic, such as code across functions or across files. Reliability also drops for very long prompts (signal dilution) and for ultra-short snippets (insufficient context).

\section{Threats TO Validity}

\mypara{Internal Validity}
A primary threat stems from \emph{fine-tuning the base generator}.  
The off-the-shelf checkpoints produced many compilation and runtime errors, masking the semantic errors that SemGuard is designed to address and exhibiting large performance variance across trials.  
We therefore fine-tuned each backbone on a small set of high-quality solutions to suppress low-level faults, stabilize the baseline, and create headroom for measuring semantic improvements.  
Although this tuning may bias absolute pass rates, it affects all decoding strategies equally and thus preserves the relative comparison. Our evaluator assumes fragment-level semantics, focusing on local logical deviations, while non-local logic (e.g., across files) is a direction we plan to explore in the future.
Randomness in both generation and evaluator training constitutes a second threat.  
To ensure reproducibility we fix all pseudo-random seeds, run every experiment three times, and report the mean; the standard deviation is below two percentage points for all metrics. 

\mypara{External Validity}
The generalizability of our results depends on the benchmark suite.  
To reduce selection bias we evaluate on two widely used Python datasets,
\textit{MBPP}~\cite{austin2021programsynthesis} and
\textit{LiveCodeBench}~\cite{jain2024livecodebench}, that differ in
problem length and difficulty.
For \textit{LiveCodeBench} we restrict ourselves to tasks published between
\mbox{1~July~2024} and \mbox{1~April~2025}, i.e.\ well after the
training–data cut-off of all tested LLMs, so that the evaluation probes
\emph{generalization} rather than memorization.  
To test language transfer we additionally run on the Java corpus
\textit{SemDiff-Java}.  
The consistent gains observed across three datasets and two programming
languages suggest that our conclusions are unlikely to be
dataset-specific, though future work should confirm them on other
domains and languages.


\section{Related Work}

\subsection{Decoding Strategies for Code LLMs}

Early decoding schemes for code LLMs—temperature~\cite{caccialanguage}, Top-$k$~\cite{fan2018hierarchical}, and Top-$p$~\cite{holtzmancurious} sampling—aim to trade off diversity against syntactic plausibility, yet they provide no functional feedback.  Subsequent work injects information signals from execution or analysis: PG-TD~\cite{zhangplanning} explores the beam space with test-guided Monte Carlo Tree Search; AdapT~\cite{zhu2024hot} dynamically tunes the sampling temperature; MBR-EXEC~\cite{shi2022natural} selects candidates under minimum Bayes risk using execution traces; MGD~\cite{agrawal2023guiding} and CodeGuard\,+~\cite{fu2024constrained} incorporate lightweight static analyses to enforce typing or security rules.  These methods markedly reduce syntax and runtime faults, but they remain weak on \emph{semantic} errors that do not manifest in compilation or unit-test failures.  ROCODE~\cite{jiang2024rocode} pushes this line further by backtracking on failing prefixes and currently reports state-of-the-art pass rates, yet it still relies on external tests and entropy heuristics to approximate semantic correctness.  In contrast, our work constructs \textit{SemDiff}, the first corpus that labels the exact line of semantic deviation, and trains a small evaluator that is embedded into the decoder, enabling real-time, test-free intervention. 

\subsection{Model Collaboration}

Recent work has increasingly explored \emph{multi-model collaboration} in code generation.
JumpCoder~\cite{chen2024jumpcoder} uses a dual-decoder architecture: a “planner” sketches the high-level scaffold, while a “filler” model completes the lower-level details.
Similarly, PToco~\cite{bian2025ptoco} samples two outputs from the same backbone and uses a prefix-grouping heuristic to select token spans where both agree, thus smoothing inconsistencies and improving pass rates.
Other frameworks, such as PairCoder~\cite{zhang2024pair}, employ two LLMs in which one generates a code draft and the other optimizes or corrects it; Xia et al.~\cite{xia2024unlocking} further reduce inference time by letting a smaller model draft and a larger model refine. IRCoCo~\cite{li2024ircoco} guides generation through collaboration between a generator and a reward evaluator, while RankEF~\cite{sun2024sifting} combines a generator with a ranker that leverages execution feedback to select correct candidates.
Despite architectural differences, prior methods share a post-hoc pattern: one model (or a symmetric pair) edits another’s output after generation, aiming mainly at \emph{syntactic} correctness or stability rather than proactive \emph{semantic} checking.
Our approach differs in two ways: (i) a \textbf{heterogeneous collaborator}—a lightweight \emph{semantic evaluator} trained on line-level deviations—and (ii) \emph{in-decoding} collaboration: the evaluator provides real-time feedback; when a line is flagged, the generator rolls back and continues with mild penalties, preventing semantic errors from propagating

\section{Conclusion and Future Work}

We presented SemGuard, a lightweight, fragment-level semantic evaluator integrated into decoding that checks partial code and backtracks upon predicted semantic deviations, curbing error propagation. Across multiple benchmarks (MBPP, LiveCodeBench, and SemDiff/Java), SemGuard consistently reduces semantic errors and outperforms prior decoding methods at comparable cost.
Future work will extend SemGuard beyond snippet-level to multi-file/framework settings and explore joint generator–evaluator training to strengthen semantic alignment while further reducing false positives and latency.

\section{ACKNOWLEDGMENTS}

This work was supported by the National Natural Science Foundation of China (NSFC) under Grant Nos. 62192731, 62192730, and 61602286.

\balance
\bibliographystyle{IEEEtran}
\bibliography{main}
\vspace{12pt}

\end{document}